\documentclass[aps,prb,preprint,groupedaddress]{revtex4}
\usepackage{graphicx}% Include figure files
\usepackage{dcolumn}% Align table columns on decimal point
\usepackage{bm}% bold math
\usepackage{amsmath,amssymb}

\def\cH{{\mathcal H}}

\def\vS{\mbox{\boldmath $S$}}

\begin{document}

%\preprint{}

\title{Quantum Phase Transitions of the Distorted Diamond Spin Chain}



\author{Tomosuke \textsc{Zenda}$^{1}$, Yuta \textsc{Tachibana}$^{1}$, 
Yuki \textsc{Ueno}$^1$, \\
Kiyomi \textsc{Okamoto}$^1$ and T\^oru \textsc{Sakai}$^{1,2}$}
\affiliation{
$^{1}$Graduate School of Material Science, University of Hyogo, Hyogo 678-1297, Japan \\
$^{2}$National Institutes for Quantum and Radiological Science and Technology (QST), SPring-8, Hyogo 679-5148, Japan
}


\date{Received September 8, 2019}

\begin{abstract}

The frustrated quantum spin system on the distorted diamond chain lattice 
suitable for the alumoklyuchevskite 
is investigated using the numerical diagonalization of finite-size clusters 
and the level spectroscopy analysis. 
It is found that this model exhibits three quantum phases; 
the ferrimagnetic phase, the spin gap one, and the gapless 
Tomonaga-Luttinger liquid depending on the exchange coupling parameters. 
The ground state phase diagram is presented.
\end{abstract}

%\pacs{75.10.Jm,  75.30.Kz, 75.40.Cx, 75.45.+j}

\maketitle

%\section{}
%\subsection{}
%\subsubsection{}

%***********************
%***********************
\section{Introduction}

Frustrated quantum spin systems have attracted a lot of interest in the field of strongly correlated electron systems. 
The $S=1/2$ distorted diamond spin chain is one of strongly frustrated quantum spin systems. 
It was proposed as a good theoretical model of the compound Cu$_3$(CO$_3$)$_2$(OH)$_2$, called azurite\cite{kikuchi}. 
The previous theoretical work\cite{okamoto1} using the perturbation analysis, 
the numerical exact diagonalization of finite clusters, and the level spectroscopy method, 
indicated that the system exhibits various quantum phases; the spin gap phase, 
the ferromagnetic one, and gapless Tomonaga-Luttinger liquid (TLL) one in the ground state, depending on the exchange coupling parameters.
Recently another candidate material of the distorted diamond spin chain was discovered. 
It is the compound  K$_3$Cu$_3$AlO$_2$(SO$_4$)$_4$, called alumoklyuchevskite\cite{fujihara,morita,fujihala2}.
This material has a different structure of the distortion from azurite.
Thus it would be useful to investigate the suitable theoretical model for alumoklyuchevskite.
In this paper, the $S=1/2$ distorted diamond spin chain model suitable for
 alumoklyuchevskite is studied by the numerical exact diagonalization of finite-size clusters
  and the level spectroscopy analysis.

\section{Model}

\begin{figure}[h]
%\centerline{\includegraphics[scale=0.5]{model-0.eps}}
\begin{center}
           \scalebox{0.5}{\includegraphics{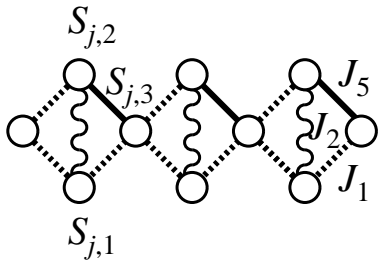}}
\end{center}
\caption{The model of the $S=1/2$ distorted diamond spin chain. 
}
\label{model}
\end{figure}

We investigate the model described by the Hamiltonian
\begin{eqnarray}
    &&\cH = \cH_0 + \cH_1 \\
    &&\cH_0 = \sum_j^L \left\{ J_2 \vS_{j,1} \cdot \vS_{j,2} + J_5 \vS_{j,2} \cdot \vS_{j,3}
                           + J_1 \vS_{j,3} \cdot \vS_{j,1} \right\} \\
    &&\cH_1 = J_1 \sum_j^L \left\{ \vS_{j,3} \cdot \vS_{j+1,2} + \vS_{j,3} \cdot \vS_{j+1,1} \right\}
\end{eqnarray}
where $\vS_{j,i}$ is the spin-1/2 operator, $J_1$, $J_2$, and $J_5$ are the 
coupling constants of the exchange interactions. 
The schematic picture of the model is shown in Fig. \ref{model}.

For alumoklyuchevskite, it is thought that the interactions corresponding to four sides of diamond
differ from one another. Since such a model, however, has many parameters, we use a simplified model
sketched in Fig.\ref{model}.
When $J_5$ is much larger than other couplings, the spins coupled by $J_5$ are going to form a singlet pairs,
which make $\vS_{j,1}$ spins nearly free. 
If the direct or effective interactions between $\vS_{j,1}$ spins are antiferromagnetic,
the ground state will be the TLL state.
This is the essential mechanism for the TLL ground state observed in
almoklyuchevskite.
On the other hand, when $J_2$ is much larger than other couplings,
singlet pairs locate at the $J_2$ bonds, which yields nearly free $\vS_{j,3}$ spins.
If the direct or effective interactions between $\vS_{j,3}$ spins are antiferromagnetic,
the ground state will be the TLL state,
which is nothing but the essential mechanism for the TLL ground state of azurite.
Thus our model is a minimal model describing both TLL ground states of alumoklyuchevskite and azurite.
We note that the direct interactions between nearly free spins are very important to explain experimental results both of almoklyuchevskite and azurite \cite{jeschke}.

For $L$-unit systems, 
the lowest energy of ${\cal H}_0$ in the subspace where 
$\sum _j S_j^z=M$, is denoted as $E(L,M)$. 
The reduced magnetization $m$ is defined as $m=M/M_{\rm s}$, 
where $M_{\rm s}$ denotes the saturation of the magnetization, 
namely $M_{\rm s}=2L/3$ for this system. 
$E(L,M)$ is calculated by the Lanczos algorithm under the 
periodic boundary condition ($\vS_{L+1,i}=\vS_{1,i}$) 
for $L=$4, 6 and 8.

\section{Ground state phase diagram}

We consider the ground state phase diagram of the model (1). 
Since the three different exchange interactions $J_1$, $J_2$ 
and $J_5$, we fix $J_1=1$ and vary $J_2$ and $J_5$ 
in this paper. 
On the analogy of the azurite-type model, 
the ferrimagnetic, the spin gap and the gapless TLL 
phases are expected to appear. 

\subsection{Ferrimagnetic phase}

\begin{figure}[tbh]
\bigskip
\centerline{\includegraphics[scale=0.35]{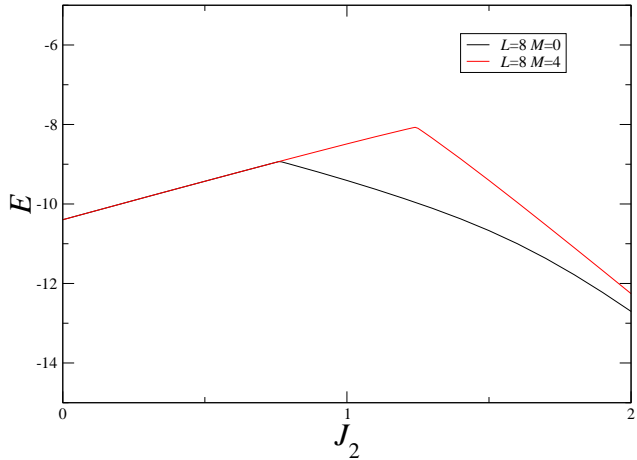}}
\caption{
$J_2$ dependence of $E(L=8,M=0)$ (black line) and $E(L=8,M=4)$ (red line) 
for $J_5=0.5$. 
The phase boundary between the ferrimagnetic and singlet 
phases can be detected as the intersection of two energy levels. 
}
\label{ferri}
\end{figure}

The ferrimagnetic phase is easily distinguished from other phases. 
In this phase the finite magnetization $m=1/3$ appears in the 
ground state. 
When $J_5=0.5$ is fixed, 
the $J_2$ dependence of the lowest energies with $M=0$ and $M=4$ for $L=8$ 
are shown in Fig. \ref{ferri}. 
The phase boundary between the ferrimagnetic and singlet 
phases can be detected as the intersection of two energy levels. 
Since the phase boundary is almost independent of the system size, 
the phase boundary is estimated from the result for $L=8$. 

\subsection{Spin gap and TLL phases}

\begin{figure}[tbh]
\centerline{\includegraphics[scale=0.35]{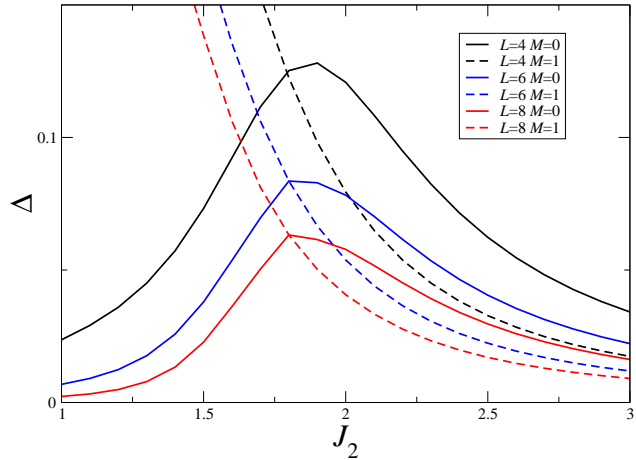}}
\caption{
$J_2$ dependences of $\Delta(L,M=0)$ and $\Delta(L,M=1)$ with $J_5=0.5$ 
for $L=4$ (black lines), 6 (blue lines) and 8 (red lines). 
Solid and dashed lines correspond to the $\Delta(L,M=0)$ and $\Delta(L,M=1)$, 
respectively. 
}
\label{LS}
\end{figure}

In order to determine the phase boundary between the 
spin gap and the TLL phases, 
the level spectroscopy analysis \cite{okamoto2,nomura} is one of the best methods. 
According to this method, we should compare the excitation energies of the 
lowest singlet excitation and the lowest triplet one. 
Namely, we define two excitation energies
\begin{eqnarray}
    &&\Delta(L,M=0) = E_1(L,M=0) - E_0(L,M=0), \\
    &&\Delta(L,M=1) = E_0(L,M=1) - E_0(L,M=0),
\end{eqnarray}
where $E_0(L,M)$ and $E_1(L,M)$ are, respectively,
the lowest energy and first excited energy within the subspace of $M$ for the $L$-unit system,
The ground state is in the spin gap phase or the TLL phase
according as $\Delta(L,M=0) > \Delta(L,M=1)$ or $\Delta(L,M=0) < \Delta(L,M=1)$.
The $J_2$ dependences of $\Delta$'s with fixed $J_5=0.5$
for $L=4$, 6 and 8 are shown in Fig. \ref{LS}. 
Assuming the finite-size correction of the cross points between 
$\Delta(L,M=0)$ and $\Delta(L,M=1)$ is proportional to $1/L^2$, 
we estimate the phase boundary in the thermodynamic limit. 
This analysis indicates that the spin gap phase is 
adjacent to the ferrimagnetic phase.

\subsection{Phase diagram}

\begin{figure}[tbh]
\centerline{\includegraphics[scale=0.35]{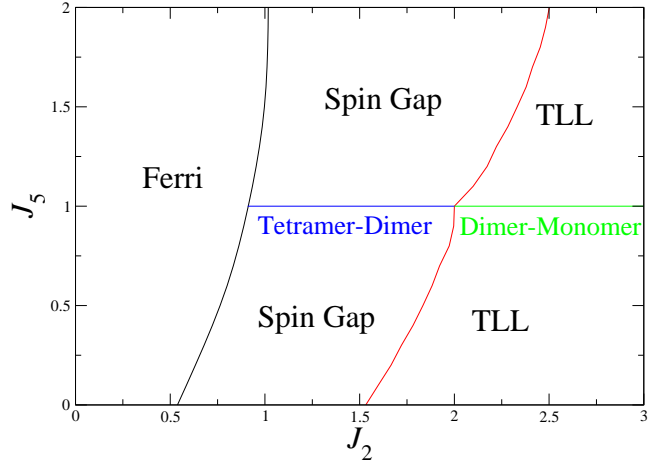}}
\caption{
Ground state phase diagram of the present model. 
It includes the ferrimagnetic, the spin gap and the TLL 
phases. 
On the line of $J_5=1$, the dimer-monomer state is the exact 
ground state for $J_2>2$ and the tetramer-dimer state is the exact ground state for $0.909 < J_2 <2$ \cite{takano}. 
}
\label{phase}
\end{figure}

According to the above analyses, the ground state phase diagram is 
obtained as shown in Fig. \ref{phase}. 
As expected, it includes the ferrimagnetic, the spin gap and the TLL phases. 
Takano {\it et al.} \cite{takano} indicated that the dimer-monomer state with high degeneracy is 
the exact ground state on the line of $J_5=1$ and $J_2>2$. 
They also found that the doubly degenerate tetramer-dimer state is the ground state on the line of $J_5=1$ and $0.909 < J_2 <2$.
Reflecting this fact, our spin gap state is also doubly degenerate which is consistent with the level spectroscopy method to
determine the boundary between the spin-gap phase and the TLL phase.

\section{Summary}

Using the numerical exact diagonalization and the level spectroscopy 
analysis, the $S=1/2$ distorted diamond spin chain suitable for the 
alumoklyuchevskite is investigated. 
The obtained ground state phase diagram includes 
the ferrimagnetic, the spin gap and the TLL 
phases.
We believe that the upper TLL state is attributed to nearly free $\vS_{j,1}$ spins (alumoklyuchevskite type),
while the lower TLL state to nearly free $\vS_{j,3}$ spins (azurite type).
More detailed analysis will be a future problem.

\section*{Acknowledgment}
This work was partly supported by JSPS KAKENHI, Grant Numbers 16K05419, 
16H01080 (J-Physics) and 18H04330 (J-Physics). 
A part of the computations was performed using 
facilities of the Supercomputer Center, 
Institute for Solid State Physics, University of Tokyo, 
and the Computer Room, Yukawa Institute for Theoretical Physics, 
Kyoto University.

\end{document}